\documentclass[conference]{IEEEtran}
\usepackage[utf8]{inputenc}
\usepackage[utf8]{inputenc}
\usepackage[english]{babel}
\usepackage[T1]{fontenc}
\usepackage{amsmath}
\usepackage{amsfonts}
\usepackage{amssymb}
\usepackage{url}
\usepackage[official]{eurosym}
\usepackage{pgfplots}
\pgfplotsset{width=7cm}
\usetikzlibrary{patterns}
\usepackage{tikz-cd}
\usetikzlibrary{shapes,arrows}
\usetikzlibrary{calc,fit,trees,positioning,arrows,chains,shapes.geometric,shapes}
\usetikzlibrary{shapes,arrows}
\author{\IEEEauthorblockN{Ricardo Morla}
\IEEEauthorblockA{ricardo.morla@fe.up.pt\\ INESC TEC and University of Porto\\Porto, Portugal
}}
\title{An initial study of the effect of pipelining \\in hiding HTTP/2.0 response sizes}
\date{May-July 2016}
\renewcommand{\footnotesize}{\scriptsize}

\begin{document}

\maketitle
\thispagestyle{plain}
\pagestyle{plain}

\begin{abstract}
HTTP response size is a well-known side channel attack. With the deployment of HTTP/2.0, response size attacks are generally dismissed with the argument that pipelining and response multiplexing prevent eavesdroppers from finding out response sizes. Yet the extent to which pipelining and response multiplexing actually hide HTTP response sizes has not been adequately investigated. In this paper we set out to help understand the effect of pipelining in hiding the size of web objects on the Internet. We conduct an experiment that provides browser-side HTTP response sizes and network-captured TLS record sizes and show how the model that we propose for estimating response sizes from TLS record sizes improves response matching and attack performance. In this process we gather evidence on how different implementations of HTTP/2.0 web servers generate different side-channel information and the limited amount of pipelining and response multiplexing used on the Internet today. 
\end{abstract}

\section{Introduction}
	HTTP with TLS encryption prevents attacks that inspect HTTP payload and signaling. HTTP response size analysis is a well-known side-channel attack~\cite{hintz_fingerprinting_2002} that overcomes encrypted payload inspection by using eavesdropped sizes of web objects to identify web applications.  Up to HTTP/1.1, the web client typically waits for the response to the current HTTP request before issuing the next request, making it straightforward to find the size of HTTP responses by tapping into the TCP/IP connection. With the deployment of HTTP/2.0~\cite{varvello_is_2016} and its pipelining and multiplexing mechanisms, most authors assume HTTP response size analysis attacks can be prevented. With request pipelining, clients no longer need to wait for the response to the current HTTP request to issue the next request. With response multiplexing, servers no longer need to wait for the current response to be finished to send the next response. Distinguishing web object sizes by eavesdropping TLS record sizes should thus be unfeasible.

Yet the extent to which pipelining and multiplexing actually hide HTTP response sizes on the Web has not been adequately investigated. The fact that these mechanisms exist and are deployed does not mean that they are used and that they are effective in hiding response sizes. Web content from a web page is often not pulled from the server at once and, if it is, HTTP signaling information may leak through TLS to help the attacker. This means that the privacy of regular web site users who will not have any particular reason for using anonymity tools like Tor~\cite{he_novel_2014} may be more at risk than what is believed and that the transition to HTTP/2.0 alone may not fully prevent this risk. This is especially relevant for the growing amount of traffic that goes through proxies and content delivery networks that share IP addresses between applications and for which a simple IP database lookup would not suffice for identifying web sites and applications.

In this paper we set out to help understand the effect of pipelining in hiding the size of web objects. The basis for this work is a set of HTTPS requests to the first 5k of Alexa's top 1M web site pages \cite{durumeric_analysis_2013}. For each web site page we collect a tcpdump capture for all incoming and outgoing traffic through the client's Ethernet interface and the HAR~\cite{odvarko2012http} log of HTTP requests and their responses as recorded by the browser. We use tcpdump 4.5.1, tshark 14.04, and Firefox 45.0.1 on Ubuntu 14.04 desktop Linux with Selenium python webdriver 2.53.1 and haralyzer 1.4.10 suites. We then extract TLS record sizes and estimate HTTP response sizes assuming 1) keep-alive, persistent TCP connections and 2) pipelining and multiplexing with unique TLS size response separator and distinct TLS record sizes for HTTP/2.0 response headers. Finally we match the estimated response sizes to the browser HAR response sizes and obtain HAR-TLS response size match results.

In the rest of this paper we start by characterizing the HAR data set that resulted from our Alexa's web site page requests. Then we describe in more detail our approach to estimate HTTP/2.0 response sizes from TLS records. HTTP/2.0 web servers pack HTTP responses differently and choose to focus on the type of web servers with most TCP connections. For these web servers we characterize our estimated HTTP/2.0 response sizes, describe the HAR-TLS estimated response size match results, and show the impact of pipelining in these results.

\section{Collected data set}

Each of the 5k Alexa's web site names was prepended with the "https://" prefix and this URL loaded into Firefox using Selenium's webdriver suite. The browser was only able to open 72\% of the web site URLs. The remaining web sites were either not active or not responding to HTTPS requests. Each web page site that responded with HTML and browser-side scripts triggers the browser to issue additional requests that are then rendered on the browser along side the main web site content. We analyzed HAR logs and found that we had recorded 328k responses: 51\% HTTPS and 18\% HTTP/2.0. We also found that less than 17\% web sites received only HTTPS additional responses, which means that for 83\% web sites part of the content that is rendered to the user is vulnerable to payload inspection attacks. Less then 7\% web sites received only HTTP/2.0 additional responses, which means that for 10\% we sites part of the content that is rendered to the user is vulnerable to directly observable response size side-channel attacks. 

These numbers tell us that we should be concerned firstly about the 83\% web sites for which part of their rendered content is vulnerable to payload inspection, and secondly about the 10\% web sites for which part of their rendered content is vulnerable to directly observable response size side-channel attacks. In this paper we assume that the trend of increased use of HTTPS and HTTP/2.0 will continue~\cite{varvello_is_2016} and that this will address the two concerns above. Consequently we focus on HTTP/2.0 and its ability to deter response size side-channel attacks.

Additional web site requests in our data set are often served by the same web server. It is thus possible to characterize individual web servers according to their use of HTTPS and HTTP/2.0. Here we define a web server as a 3-tuple \texttt{(IP, port number, protocol)} where \texttt{IP} is the web server IP address, \texttt{port} is the TCP/IP port number  (80 or 443), and \texttt{protocol} is either HTTP/1.1 or HTTP/2.0. We have responses from 20k web servers, 46\% of which are \texttt{(*, 443, HTTP/1.1)}  and 6\%  \texttt{(*, 443, HTTP/2.0)}. The top 10 web server response is at 5\% for responses from \texttt{(*, 443, HTTP/1.1)} web servers and at 34\% for \texttt{(*, 443, HTTP/2.0)} web servers. Top 7 HTTP/2.0 web server names by HTTP request count as reported by the web servers and logged in HAR is 'sffe': 16\%, '': 15\%, 'cloudflare-nginx': 14\%, 'cafe': 14\%, 'Golfe2': 8\%, 'adclick\_server': 6\%, 'nginx': 5\%.

\section{Estimating HTTP response sizes }
Our approach to estimate HTTP response sizes from TLS records has two parts. In the first part we address the feature in HTTP/1.1 and beyond that allows a TCP connection to be reused for multiple requests. We segment the TLS records of each TCP connection into sets of HTTP request-response sequences that are contained entirely in each segment. Our assumption in segmenting TCP connections is that the TLS records of ready-to-send HTTP responses are sent back-to-back. We look for gaps in the sequence of timestamps of these records and use these gaps to segment the TCP connections.  In HTTP/1.1 a gap will exist between the timestamp of the last TLS record of the current response and the timestampt of the first TLS record of the next response. This gap is at least one round trip time, as the client has to wait until the the current response is completely received before sending the next request. More generically in order to include pipelining and response multiplexing are included, we take a gap in the back-to-back TLS record sequences from the server as indication that all requests sent by from the client and received by the server up to that point have been served. We declare a gap in the TCP connection when the difference between the timestamps of consecutive TLS records from the server is 1) larger than 0.5 seconds or 2) larger than 20 times the average back-to-back response gap from the server, normalized to a 1500 byte TLS record size.

In the second part of our approach we analyze the sets of HTTP request-response of the segmented TCP connections to compute response sizes. The basic keep-alive model does not consider pipelining and multiplexing and simply takes the sum of all TLS record sizes from the server on each segment of the TCP connection as an estimate of HTTP response size.  model that considers pipelining and multiplexing to estimate HTTP response sizes relies on our intuition on the following three side-channel information that leak from TLS. 1) HTTP/2.0 signaling is visible through small-sized TLS records (less than 60 bytes) that indicate the beginning of an HTTPS connection or the beginning or end of an HTTP response. 2) Large HTTP responses generate back-to-back TLS records with either network ($\sim$ 1.5 kB) or TLS ($\sim$ 16 kB) MTUs. 3) Request and response headers are sent in their own TLS records, typically yielding record sizes smaller than the MTU. We use this information to estimate the start and finish times for HTTP requests and responses. With start and finish times it is possible to identify segments in the TLS record size sequences that contain entire responses that are not multiplexed and others that are entirely or only partially multiplexed. We sum all TLS record sizes to get an estimate of non-multiplexed responses and ignore multiplexed responses.

What are the specific TLS record sizes that can be used to determine start and finish times for requests and responses? After manually inspecting the sequence of TLS record sizes of the connections for the first few web site pages on Alexa's top web sites it is clear that the HTTP/2.0 data and signaling is encapsulated differently for different web server implementations. We would like to group web servers in types according to how they encapsulate HTTP/2.0 data and signaling. In order to do so, we use the first one to three small (<100 bytes) TLS record sizes sent by the server at the beginning of each HTTP/2.0 TCP connection and call this the web server type sequence. We found 21k HTTP/2.0 TCP connections and 1252 unique HTTP/2.0 web server IP addresses in our dataset. TLS handshaking has an application-layer protocol negotiation (ALNP) extension in which the type of application data is sent in clear text. We use this to establish if a TCP connection is being used for HTTP/2.0 or not. We define HTTP/2.0 web servers as a web server for which we can observe at least one HTTP/2.0 TCP connection.

The first observation regarding types of web servers is that all the connections to a given web server have the same initial type sequence. The second observation is that the top five type sequences account for more than 99\% of the HTTP/2.0 TCP connections and over 89\% of the HTTP/2.0 web servers. Figure \ref{webserver-types} shows the percentage of TCP connections and web servers for the first five type sequences. For each web server IP address found in the packet capture logs we find the server reported names on the HAR logs. We also report this information in Figure \ref{webserver-types} together with a whois lookup for the web servers IP addresses. 

\begin{figure}[]
\begin{center}
\small
\begin{tikzpicture}[scale=0.9]
\begin{axis}[
    ybar,
    ytick={0,20,40,60,80,100},
    ymax = 100,
    ymajorgrids=true,
    xmajorgrids=true,
    enlargelimits=0.10,
    legend style={at={(1.33,0.3)},
      anchor=north,legend columns=1, font=\footnotesize},
    ylabel={Percentage},
    symbolic x coords={Type 1,Type 2,Type 3, Type 4, Type 5},
    xtick=data,
    %nodes near coords,
    nodes near coords align={vertical},
    ]
\addplot [black, fill=gray] coordinates {(Type 1,69.5) (Type 2,16.2) (Type 3,8.4) (Type 4,2.9) (Type 5,2.9)};
\addplot [black, fill=lightgray] coordinates {(Type 1,18.0) (Type 2,1.8) (Type 3,68.0) (Type 4,1.0) (Type 5,0.9)};
\legend {TCP Connections,Web servers}
\end{axis}
\node [font=\small, anchor=west] (top)    at (5.5,4) {Type 1: [51,37,33]};
\node [font=\small, anchor=west] (middle) at (5.5,3.5)  {Type 2: [76,33,37]};
\node [font=\small, anchor=west] (middle) at (5.5,3)  {Type 3: [64,33]};
\node [font=\small, anchor=west] (middle) at (5.5,2.5)  {Type 4: [39,33,37]};
\node [font=\small, anchor=west] (middle) at (5.5,2)  {Type 5: [76,33,41]};
\end{tikzpicture}
\vspace{1pt}

\small
  \begin{tabular}{ | c | c | c | }
    \hline
     & whois & Reported names \\ \hline \hline
	Type 1 & Google & "ssfe", "", "HTTP server (unknown)", \\ & & "safe", "Golfe2", "cafe", "xbfe", \\ & & "GSE", "gws", ... \\ \hline
    Type 2 & Facebook & "", "proxygen" \\ \hline
    Type 3 & Cloudflare, & "cloudfare-nginx", "nginx", "", \\ & Taobao, ... & "Tengine", ... \\ \hline
    Type 4 & Twitter & "tsa\_f", "" \\ \hline
    Type 5 & Facebook & "", "proxygen" \\ \hline
  \end{tabular}
  \vspace{5pt}

\caption{Percentage of HTTP/2.0 TCP connections and web servers per type (top). Web server IP address whois lookup and web server names as reported to by server to browser (bottom).}
\label{webserver-types}
\end{center}
\end{figure}

In this paper we focus on Type 1 servers that are responsible for most of the HTTP/2.0 TCP connections in our data set. Manual inspection of data from these servers allowed us to define the following assumptions for our model. 

\begin{itemize}
\item HTTP request and response headers are sent in TLS records with sizes ranging from 70 to 350 bytes.
\item 41 byte-length TLS records from the server indicate that a response has finished.
\end{itemize}

The output of the estimation process is a list of sizes of estimated responses from each web server per web page site visited.

\section{Methodology}

The algorithm that allows us to determined the number of matched bytes and requests for each web server and each visited web page site is as follows. First we define a matrix with the byte difference between every estimated response and every HAR response. We attempt to match smaller positive entries in the matrix first. By requiring a positive entry we are making sure that the estimated response is larger than the HAR response. HAR response size is the size of the compressed HTTP response, which should be equal to the response size estimated from the TLS records or smaller if any padding occurs at the TLS layer. For the smallest positive entry we find at each iteration, we declare a match between the estimated response corresponding to the entry's row and to the HAR response corresponding to the entry's column, push that match into a list, and set all entries in that row and column of the matrix to 'NAN'. After the last iteration all entries in the matrix should be 'NAN'. We then remove matches whose difference between estimated response size and HAR response size are 1) 15\% or larger or 2) larger than 200 bytes. 

The number of matched bytes and requests per web server and per  web page site is then used to compute two performance metrics. These performance metrics are the ratio of matched bytes to bytes in the HAR logs and the ratio of matched responses to number of responses in the HAR logs. We measure the performance of our estimation over the set of requests that each HTTP/2.0 web server responds to when each web site page is visited. We compare the performance of our type 1 server model with that of the simple keep-alive model using empirical cumulative distributions over the set of $<$web server, web page site$>$ tuples.

\section{Results}

We obtained over 30k estimated HTTP/2.0 responses from type 1 servers with an average of 2 HTTP/2.0 requests per TCP connection. Over 9k out of the 30k estimated responses were pipelined and not multiplexed. We also found 1.5k sets of multiplexed responses corresponding to 4k out of 30k HTTP/2.0 responses.

\begin{figure}[h!]
\begin{center}
\includegraphics[width=0.45 \textwidth]{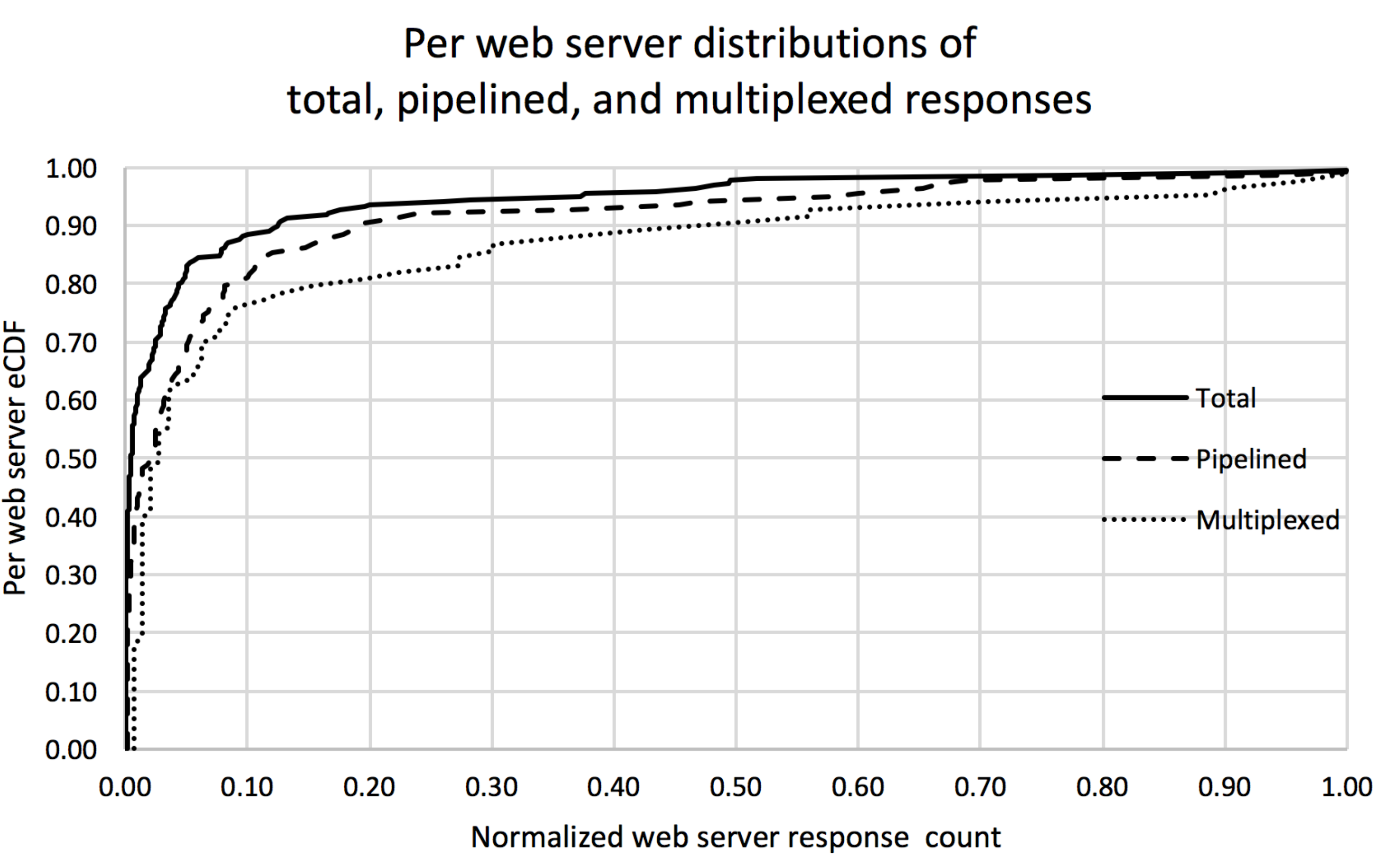} 
\end{center}
\caption{Comparing total, pipelined, and multiplexed per web server response distribution.}
\label{per-ws-pipel-mux}
\end{figure}

We observed 218 type 1 web servers in our estimated HTTP/2.0 responses, out of which 140 provided more than one pipelined response and 83 at least one set of multiplexed responses. Figure \ref{per-ws-pipel-mux} shows the per web server distribution of responses normalized to  the maximum response count for total, pipelined, and multiplexed responses. The similarity in the shape of these distributions suggests that although less frequent, pipelining and multiplexing occurs in the same proportion throughout the servers.

We found 143 type 1 web servers in the HAR records of our data set. These servers generated over 16.9k HTTP2/.0 responses. Out of these 16.9k responses, we found a match for 8.8k responses with the basic keep-alive model and for 12.0k with type 1 server model. This corresponds to an overall matching improvement of 3.3k responses. Regarding matched bytes the results are similar: matched 240 MB with the keep-alive model and 330 MB with the type 1 server model, corresponding to a 90 MB improvement. Out of the 16.9k responses, we have an improvement roughly from 50\% to 70\% matched responses with less than 15\% and 200 bytes TLS overhead.

What is the impact of pipelining in these improvements? We found that 1) 3.9k type 1 matched responses were pipelined, 2) 3.7k pipelined responses had a match with type 1 model and no match using the simple keep-alive model. This means that 94\% of the pipelined responses are new matches brought in by the type 1 server model. The impact of multiplexing is more difficult to ascertain as we do not estimate the size of the responses only how many. A statistical dependency study or a model for estimating multiplexed response sizes would be required to understand such an impact.

\begin{figure}
\begin{center}
\includegraphics[width=0.45 \textwidth]{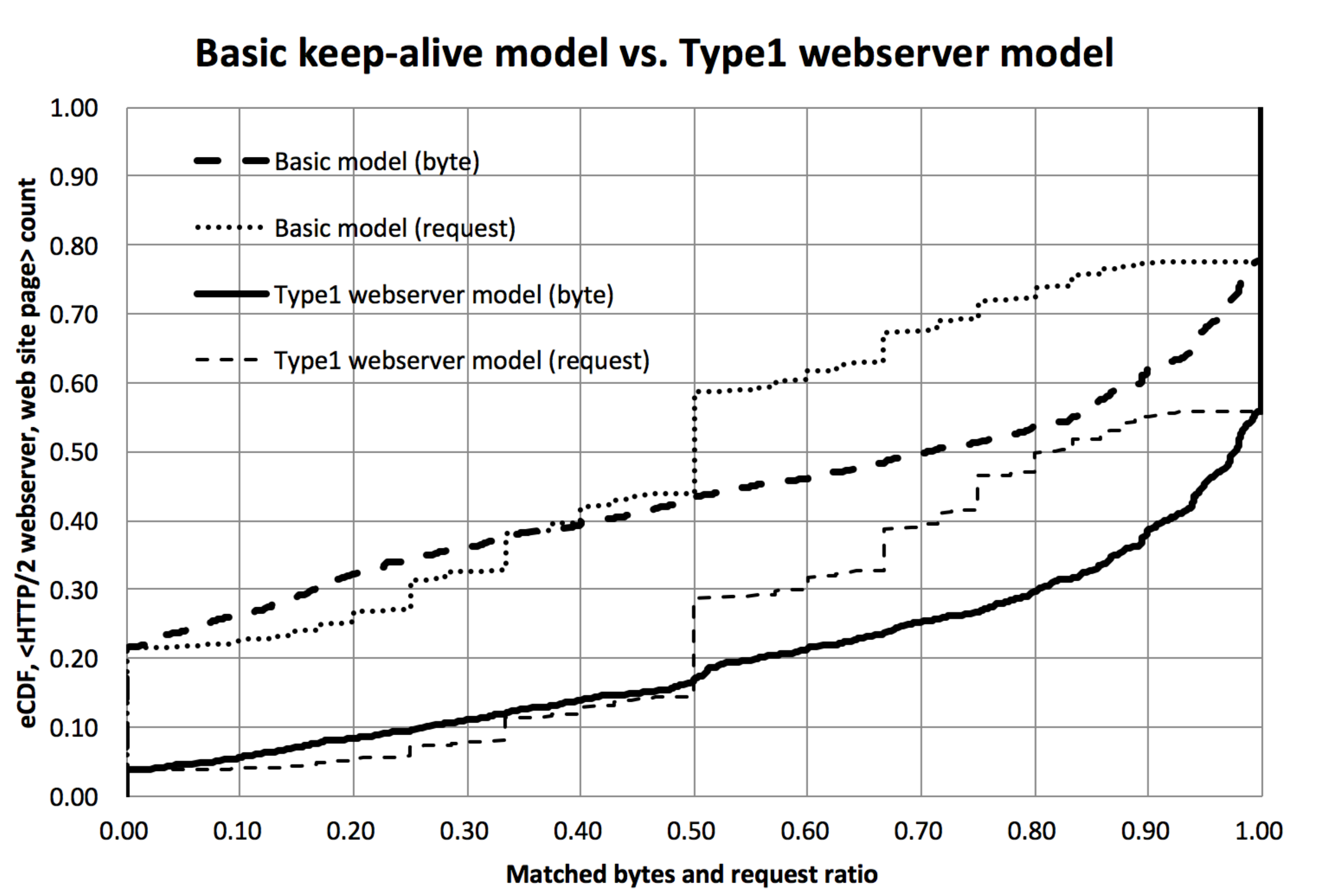} 
\end{center}
\caption{Comparison of keep-alive and type 1 server models.}
\label{results-ka-vs-type1}
\end{figure}

Figure \ref{results-ka-vs-type1} shows an overall improvement in matched bytes and requests for type 1 web server model compared to the keep-alive model. In particular we observe a 21.6\% to 3.8\% reduction of no matched bytes and no matched requests and a 22.4\% to 44.1\% increase of all matched bytes and all matched requests when comparing the type 1 web server model to the keep-alive model.

\section{Conclusion}
Our conclusions from this experiment are as follows: 1) the amount of pipelining observed is moderate ($\sim$ 30\%); 2) the amount of multiplexing is small ($\sim$ 13\%); 3) the use of TLS side-channel  markers improves response matching; and 4) almost all pipelined responses could not be matched under simple model and can be matched with improved model.

Our recommendation is to engage into effectively using pipelining and request multiplexing, by limiting the effect of TLS side-channel information and increasing response multiplexing.

The main caveats of this work is whether these results hold for more extensive web page site data set and for other types of HTTP/2.0 web sites.

\bibliographystyle{unsrt}
\bibliography{pipelining}

\begin{thebibliography}{1}

\bibitem{hintz_fingerprinting_2002}
Andrew Hintz.
\newblock Fingerprinting {Websites} {Using} {Traffic} {Analysis}.
\newblock In Roger Dingledine and Paul Syverson, editors, {\em Privacy
  {Enhancing} {Technologies}}, number 2482 in Lecture {Notes} in {Computer}
  {Science}, pages 171--178. Springer Berlin Heidelberg, April 2002.
\newblock DOI: 10.1007/3-540-36467-6\_13.

\bibitem{varvello_is_2016}
Matteo Varvello, Kyle Schomp, David Naylor, Jeremy Blackburn, Alessandro
  Finamore, and Konstantina Papagiannaki.
\newblock Is the {Web} {HTTP}/2 {Yet}?
\newblock In Thomas Karagiannis and Xenofontas Dimitropoulos, editors, {\em
  Passive and {Active} {Measurement}}, number 9631 in Lecture {Notes} in
  {Computer} {Science}, pages 218--232. Springer International Publishing,
  March 2016.
\newblock DOI: 10.1007/978-3-319-30505-9\_17.

\bibitem{he_novel_2014}
Gaofeng He, Ming Yang, Xiaodan Gu, Junzhou Luo, and Yuanyuan Ma.
\newblock A novel active website fingerprinting attack against {Tor} anonymous
  system.
\newblock In {\em Proceedings of the 2014 {IEEE} 18th {International}
  {Conference} on {Computer} {Supported} {Cooperative} {Work} in {Design}
  ({CSCWD})}, pages 112--117, May 2014.

\bibitem{durumeric_analysis_2013}
Zakir Durumeric, James Kasten, Michael Bailey, and J.~Alex Halderman.
\newblock Analysis of the {HTTPS} {Certificate} {Ecosystem}.
\newblock In {\em Proceedings of the 2013 {Conference} on {Internet}
  {Measurement} {Conference}}, {IMC} '13, pages 291--304, New York, NY, USA,
  2013. ACM.

\bibitem{odvarko2012http}
Jan Odvarko, Arvind Jain, and Andy Davies.
\newblock {HTTP Archive (HAR) format}.
\newblock {\em W3C draft}, 2012.

\end{thebibliography}

\end{document}